\documentclass[]{aa}                    
\bibpunct{(}{)}{;}{a}{}{,}
\usepackage[varg]{txfonts}

\usepackage{nicefrac}

\usepackage[position=top]{subcaption}
\graphicspath{{fig/}}
\begin{document}

\title{Plasma dynamics in solar macrospicules \\ from high-cadence EUV observations}
\titlerunning{Plasma dynamics in solar macrospicules}
\author{ I.P. Loboda \and S.A. Bogachev }
\institute{P.N. Lebedev Physical Institute of the Russian Academy of Sciences,
           53 Leninskij prospekt, 119991 Moscow, Russia\\
           \email{loboda@sci.lebedev.ru} }
\date{Received date / Accepted date }           
           
\abstract{
Macrospicules are relatively large spicule-like formations found mainly over the polar coronal holes when observing in the transition region spectral lines.
In this study, we took advantage of the two short series of observations in the \ion{He}{ii} 304 \AA\ line obtained by the TESIS solar observatory with a cadence of up to 3.5~s to study the dynamics of macrospicules in unprecedented detail.
We used a one-dimensional hydrodynamic method based on the assumption of their axial symmetry and on a simple radiative transfer model to reconstruct the evolution of the internal velocity field of 18 macrospicules from this dataset.
Besides the internal dynamics, we studied the motion of the apparent end points of the same 18 macrospicules and found 15 of them to follow parabolic trajectories with high precision which correspond closely to the obtained velocity fields.
We found that in a clear, unperturbed case these macrospicules move with a constant deceleration inconsistent with a purely ballistic motion and have roughly the same velocity along their entire axis, with the obtained decelerations typically ranging from 160 to 230~m~s$^{-2}$, and initial velocities from 80 to 130~km~s$^{-1}$.
We also found a propagating acoustic wave for one of the macrospicules and a clear linear correlation between the initial velocities of the macrospicules and their decelerations, which indicates that they may be driven by magneto-acoustic shocks.
Finally, we inverted our previous method by taking velocities from the parabolic fits to give rough estimates of the percentage of mass lost by 12 of the macrospicules.
We found that typically from 10 to 30\% of their observed mass fades out of the line (presumably being heated to higher coronal temperatures) with three exceptions of 50\% and one of 80\%. 
}
\maketitle

\section{Introduction}

Discovered over 40 years ago, macrospicules still remain one of the most enigmatic phenomena in the solar atmosphere.
They were first identified in the \ion{He}{ii} 304~\AA\ spectroheliograms obtained during the \textit{Skylab} mission in 1973 and were so named for their morphological resemblance to the conventional H$\alpha$ spicules, while being significantly larger in size and longer-lived \citep{1975ApJ...197L.133B}.
Typically, they reach heights of 7 to 40~Mm ($10\arcsec$--$60\arcsec$) with the lifetimes ranging from 3 to 45~min \citep{1975ApJ...197L.133B, 1976ApJ...203..528W}, and attain maximum velocities from 70 to 140~km~s$^{-1}$ \citep{1989SoPh..119...55D, 1994ApJ...431L..59K}.
Another distinguishing feature of macrospicules is that they are predominantly observed in coronal holes \citep{1975ApJ...197L.133B, 1998ApJ...509..461W}.

Macrospicules are most often visible in the transition region spectral lines, such as \ion{He}{ii} 304~\AA\ and \ion{O}{v} 630~\AA, formed at the temperatures of $8\times 10^4$ and $2.5\times 10^5$~K, respectively.
At the same time, they are rarely seen in the hotter \ion{Ne}{vii} 465~\AA\ line and are mostly absent in \ion{Mg}{ix} 368~\AA, which implies an upper temperature limit of $5\times 10^5$--$10^6$~K \citep{1975ApJ...197L.133B, 2011A&A...532L...1M}.
These results are supported by observations in the radio wavelengths: measurements of the brightness temperature of the 4.8~GHZ emission, most probably originating from thermal bremsstrahlung, suggest that macrospicules typically consist of a cool core at (4--8)$\times 10^3 $~K surrounded by a hotter shell at (1--2)$\times 10^5$~K, which corresponds to plasma density varying from $10^{10}$~cm$^{-3}$ in the core to $2\times 10^9$~cm$^{-3}$ in the outer sheath \citep{1991ApJ...376L..25H}. 

Similar features, also referred to as macrospicules, are observed in H$\alpha$ in the quiet Sun, but whether these are the same physical phenomena as the \ion{He}{ii}, or the extreme-UV (EUV) macrospicules remains unclear \citep{1979SoPh...61..283L}.
Some authors found similarities between H$\alpha$ and EUV macrospicules striking in both their morphology and dynamics, and suggested that EUV macrospicules most likely represent a hotter shell of H$\alpha$ macrospicules \citep{1977ApJ...218..286M}.
Others pointed out that the morphology of H$\alpha$ macrospicules is often different from that observed in \ion{He}{ii} and that more than 50\% of H$\alpha$ macrospicules in polar regions do not show up in helium, while EUV macrospicules typically have faint counterparts in H$\alpha$; moreover, unlike EUV macrospicules, H$\alpha$ macrospicules are not constrained to the coronal holes \citep{1975SoPh...40...65M, 1998ApJ...509..461W}.
In addition, \citet{1999A&A...341..610G} proposed drawing a clear distinction between macrospicules and surges; though they both look like giant spicules, the latter proved to have a more complex morphology.

It is not clear whether macrospicules are just oversized spicules or whether they are produced by an essentially different mechanism.
The fact that they are mainly found in coronal holes implies that the presence of the open magnetic field lines is critical for their formation.
Numerical simulations suggest that in the open-field configuration two different types of jets can exist, one of which presumably corresponds to macrospicules, scarcely seen in H$\alpha$ because of their low density \citep{1982SoPh...81....9S}.
Although some of the macrospicules' properties observed in the EUV have been reproduced in simulations, the formation mechanism of macrospicules remains an unresolved problem \citep{1994ESASP.373..179A, 2011A&A...535A..58M}.

Some authors note that macrospicules have approximately the same velocities as the recently discovered Type II spicules (discovered by \citealp{2007PASJ...59S.655D}), and moreover, that they are both predominantly observed in the coronal holes, which indicates that these phenomena can be closely related \citep{2010ApJ...722.1644S}.
It was estimated that the amount of energy required to produce a macrospicule exceeds that required for an ordinary spicule by roughly two orders of magnitude, and therefore, macrospicules should be considered an important factor in the mass and energy balance of the corona \citep{1976ApJ...203..528W}.
Also, macrospicules have been proposed as a possible source of the fast solar wind component because of their high velocities and the fact that both these phenomena are constrained to the coronal hole regions \citep{1994A&A...281...95L, 1997SoPh..175..457P}.

For a long time, EUV macrospicules remained close to the resolution limit of the existing space-based solar observatories.
During the past decade, however, a dramatic increase in both their spatial and temporal resolution has greatly facilitated the study of these highly dynamic phenomena.
In this work, we take advantage of the high-cadence EUV observations obtained by the TESIS solar observatory to perform a detailed investigation of the axial plasma motions in solar macrospicules.
In Section~\ref{sect:data} of this paper we give a general overview of these data and describe the models and processing techniques applied. 
We then present the obtained results in Section~\ref{sect:results} and finalise with a brief discussion in Section~\ref{sect:discussion}.

\section{Data and processing} \label{sect:data}
\subsection{Observations}

For this study, we used observations in the \ion{He}{ii} 304~\AA\ line obtained by one of the two Full-disc EUV Telescopes (FET) of TESIS.
This instrument was developed at the Lebedev Institute of the Russian Academy of Sciences and launched on board the CORONAS-Photon satellite on 30 January 2009 \citep{2011SoSyR..45..162K}.
The telescope had two EUV channels centred at 171 and 304~\AA, and during the regular phase of observations it produced images of the full solar disc and the low corona with the field of view (FOV) of $1.0\degr$, angular resolution of $1.7\arcsec$, and at a cadence of 4~min \citep{2009AdSpR..43.1001K}. 

In addition, a number of series of observations were made with a much higher cadence in both 171 and 304~\AA\ channels.
With an exposure time close to 0.5~s, the cadence was mostly limited by the read-out time of the CCD and the telemetry constraints.
For this reason, the field of view of these observations was reduced, and the observations in the two channels were not simultaneous.
We analysed two of these time series in the 304~\AA\ channel pointed at the north pole of the Sun and imaging the surrounding polar coronal hole.
The first, consisting of 299 images with a FOV of $32\arcmin\times 12\arcmin$ was made on 2 October 2009 between 13:40:37 and 13:58:09~UT at a cadence of 3.5~s.
The second, composed of 489 images, was obtained on 23 November 2009 from 12:23:32 to 13:12:32~UT with a greater FOV of $34\arcmin\times 20\arcmin$, but at a lower cadence of 6.0~s.

\begin{figure}[]
        \centering
        \includegraphics{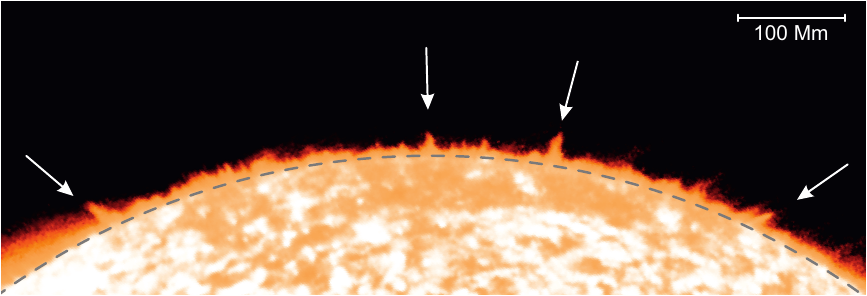}
        \caption{Snapshot of TESIS observations in the 304~\AA\ channel showing the north pole coronal hole with several macrospicules.}
        \label{fig:snapshot}
\end{figure}

In these time series we identified a total of 36 macrospicules, not all of which, however, were very distinct.
Typically, they become distinguishable from the background as separate protrusions of plasma at the heights of 10--15~Mm above the limb.
They ascend, stop, and then descend along a certain axis, showing very weak or no sideward motions.
We note that most of the macrospicules do not stretch radially from the Sun, but are inclined outward from the poles (Fig.~\ref{fig:snapshot}).

\subsection{Data reduction} \label{processing}

Given these high-cadence observations, our idea was to apply hydrodynamic methods of investigation to study the internal dynamics of macrospicules.
For this, we primarily needed to infer plasma density distribution from the intensities observed, in which endeavour we relied on the fact that macrospicules are dense enough to be optically thick in the \ion{He}{ii} 304~\AA\ line.
An implication of this is that the observed radiation should be mainly formed within a thin outer sheath of thickness $h_0$ where emission from the underlying layers of a macrospicule is absorbed by helium ions.
From the Beer-Lambert law, this thickness can be expressed as
\begin{equation}
h_0 = 1/\varepsilon \sim 1/n_{\mathrm{HeII}} \,,
\end{equation}
where $\varepsilon$ is the attenuation coefficient of the shell and $n_{\mathrm{HeII}}$ is the number density of partly ionised helium atoms.

To go further we had to assume that the emitting layer of the macrospicule is isothermal (with the temperature $T_0$) and homogeneous along the line of sight.
In this case the observed intensity is given by 
\begin{equation}
I = G(T_0)\mathrm{EM}(T_0) \sim n_{\mathrm{HeII}} n_{\mathrm{e}} h_0 \sim n_{\mathrm{e}} \,,
\end{equation}
where $G$ is the contribution function, EM is the emission measure, and $n_{\mathrm{e}}$ is the electron density in this outer sheath.
Taking into account the short lifetimes of macrospicules, the helium abundance and ionisation degree can be considered constant, and therefore plasma density $\rho$ can be taken as proportional to the electron density, {i.e.} 
\begin{equation}
\label{eq:rad_model}
\rho \sim n_{\mathrm{e}} \sim I \,. 
\end{equation}
Though this is by no means a precise model, it is the only practical way to derive the plasma density distribution within a macrospicule directly from the 304~\AA\ observations.
More accurate numerical simulations, such as those performed by \citet{2009A&A...503..663G}, which consider a realistic radiative transfer model including the scattered radiation from the Sun, can hardly be applied to process such data, making simplifications inevitable when working with optically thick plasmas.
Moreover, these simulations suggest that for the typical plasma densities in macrospicules the linear dependence given in Equation~(\ref{eq:rad_model}) is likely to be an acceptable approximation.

To investigate the dynamics of a macrospicule, we considered a one-dimensional model in which the macrospicule's mass is concentrated along its inclined axis.
Consequently, we represent the distribution of mass in terms of linear density 
\begin{equation}
\lambda = A\rho \,,
\end{equation}
where $A=A(z)$ is the macrospicule's cross-section area at a given height $z$ above the limb.
Assuming that the observed macrospicule is axially symmetric, and that for each horizontal slice the electron density in its interior remains proportional to the mean electron density in the outer sheath, we arrive at
\begin{equation}
\lambda(z) \sim \langle n_{\mathrm{e}} \rangle d^2 \sim I_{\mathrm{\Sigma}}d \,,
\label{eq:density_model}
\end{equation}
where $d = d(z)$ is the diameter and $I_{\Sigma} = I_{\Sigma}(z)$ is the net flux from a single horizontal layer of the macrospicule. 

\begin{figure}[]
        \centering
        \includegraphics{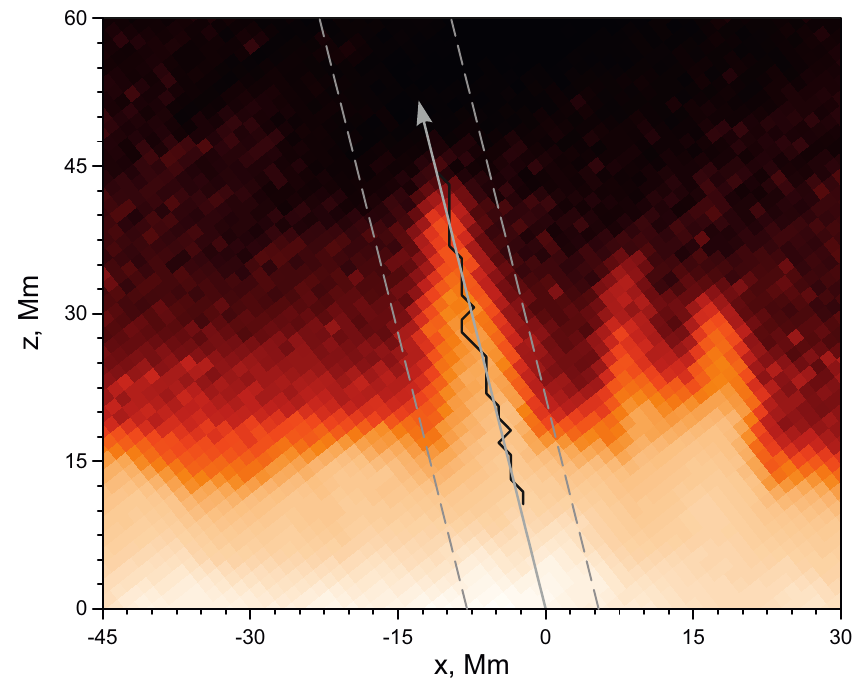}
        \caption{Maximum intensity map showing a macrospicule's spine (black polyline), axis (grey arrow), and the cut-out window (grey dashed lines). $z$ is height above the limb and $x$ is distance along the limb.} 
        \label{fig:tube}
\end{figure}

\begin{figure}[]
        \centering
        \includegraphics{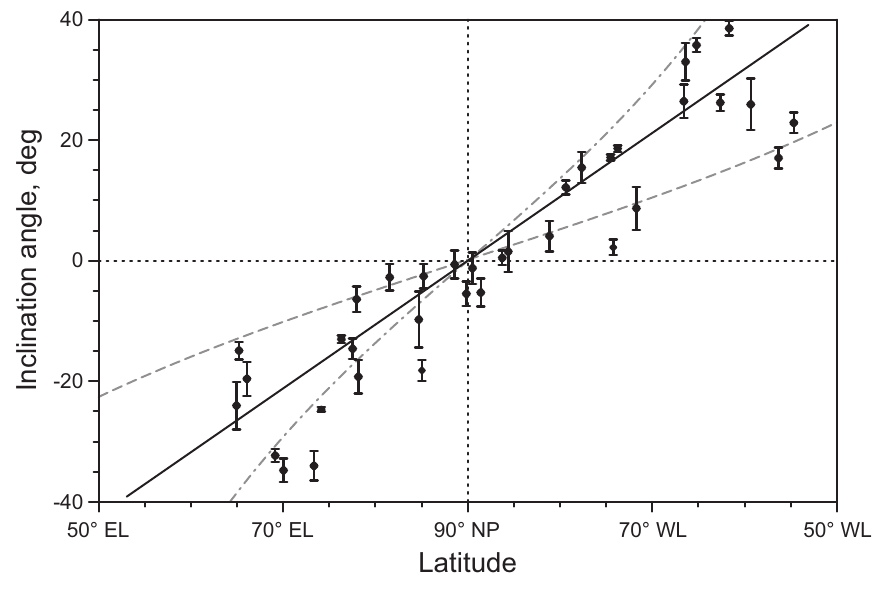}
        \caption{Apparent clockwise inclination of a macrospicule from the vertical versus the latitudinal position of its base. The linear fit to the data points is shown with the solid black line. For comparison, field line inclinations characteristic of a dipole (dashed line) and a quadrupole (dash-dotted line) are shown in grey.}
        \label{fig:inclin}
\end{figure}

To find the axis of a macrospicule, we built a map of maximum intensities defined as
\begin{equation}
M(\varphi, z) = \max_{t}I \,,
\end{equation}
where $t$ is time and $\varphi$ is the latitude (see example in Fig.~\ref{fig:tube}).
In this map, we determined the macrospicule's spine as the positions of intensity maxima at varied heights, and subsequently found its axis as a linear fit to this spine.
After performing this operation for all the macrospicules in the dataset, we established a strong linear correlation between the inclination angle of a macrospicule to the radial direction (counted clockwise) and its lattitude, with the proportionality constant $k=1.07 \pm 0.03$ and the Pearson product-moment correlation coefficient $r=0.9$ (Fig.~\ref{fig:inclin}).
Interestingly, this dependence is significantly steeper than that implied by an ideal dipole with $k=\nicefrac{1}{2}$ in the close vicinity of the poles, but flatter than that of a quadrupole, which implies $k=\nicefrac{4}{3}$ (grey lines in Fig.~\ref{fig:inclin}).
This indicates that the global structure of magnetic field lines, which guide the macrospicules, is a complex combination of several multipole modes \citep{2012ApJ...757...96D}.

To extract numerical data from the image, we manually specified a narrow cut-out window parallel to the macrospicule's axis, which was, nevertheless, wide enough to fit the macrospicule entirely at any phase of its movement (Fig.~\ref{fig:tube}).
In contrast with taking intensity values from a narrow slit, this approach is less sensitive to the possible transverse motions of a macrospicule and guarantees that all of its material will be registered \citep{2012ApJ...750...16Z}.
Following Equation~(\ref{eq:density_model}), we summed up the intensity values for each horizontal layer of the macrospicule inside this cut-out window, and multiplied this sum by the macrospicule's diameter at the same height, which was defined as the full width at half maximum of the respective horizontal intensity profile.
This means that the obtained linear density is a function of height, and therefore, the subsequent calculations will yield only the vertical component of the macrospicule's velocity field.
This approach is, nevertheless, completely equivalent to finding velocities along the macrospicule's axis, which can be easily obtained at a later stage given the inclination angle of the macrospicule.

\begin{figure}
        \centering
        \begin{subfigure}[]{\linewidth}
                \centering
                \caption{}
                \vspace{-6mm}
        \includegraphics{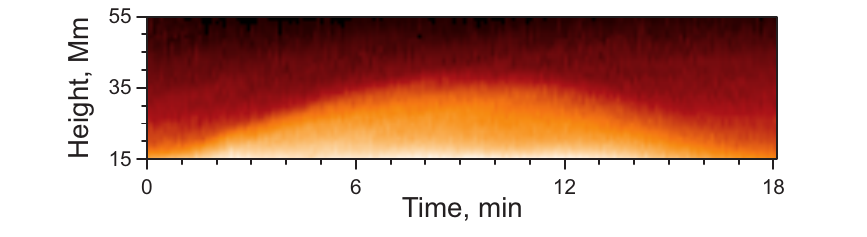}
        \label{fig:inislice}
        \end{subfigure}
                
        \begin{subfigure}[]{\linewidth}
                \centering
                \caption{} 
                \vspace{-6mm}
                \includegraphics{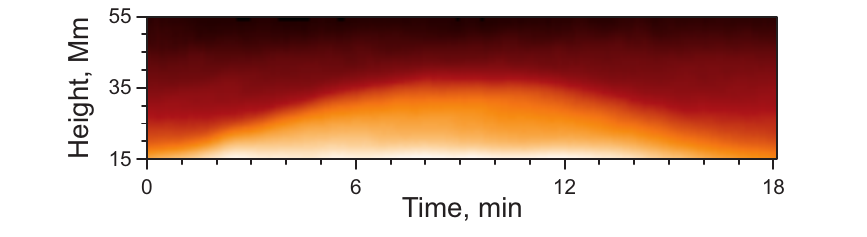}
                \label{fig:filtslice}
        \end{subfigure}

        \begin{subfigure}[]{\linewidth}
                \centering
                \caption{} 
                \vspace{-6mm}
                \includegraphics{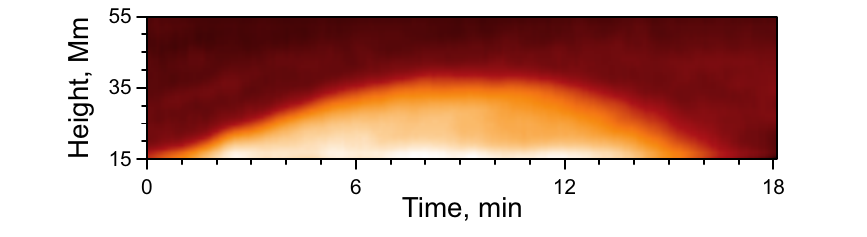}
                \label{fig:noBGslice}
        \end{subfigure}

        \begin{subfigure}[]{\linewidth}
                \centering
                \caption{} 
                \vspace{-6mm}
                \includegraphics{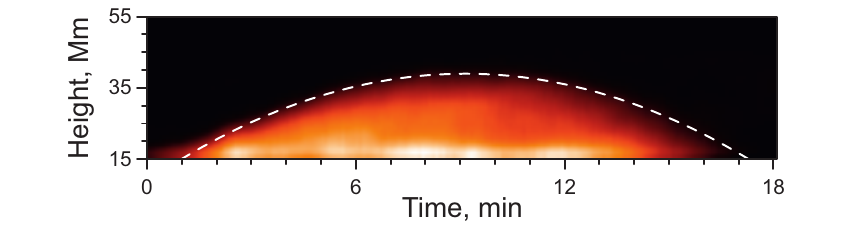}
                \label{fig:clearslice}
        \end{subfigure}
        \caption{
Space-time plots illustrating the sequential steps of the data processing pipeline: the raw density map~(\subref{fig:inislice}), the output of the low-pass filter~(\subref{fig:filtslice}), the result of background subtraction~(\subref{fig:noBGslice}), and the fully pre-processed data array~(\subref{fig:clearslice}).
The parabolic fit to the macrospicule's leading edge trajectory is indicated with the white dashed line in panel~(\subref{fig:clearslice}).
}
\label{fig:slices}
\end{figure}

Repeating the above procedure for every image in the dataset where the macrospicule was identified, we obtained a space-time density map $\lambda(z,t)$, which, within the limits of our simplified model, reflects the linear density variations in the macrospicule through space and time (Fig.~\ref{fig:inislice}).
These data, however, require some further processing before the velocity field of the macrospicule can be reconstructed.

\begin{figure}
        \centering
        \begin{subfigure}[]{\linewidth}
                \centering
                \caption{}
                \vspace{-6mm}
                \includegraphics{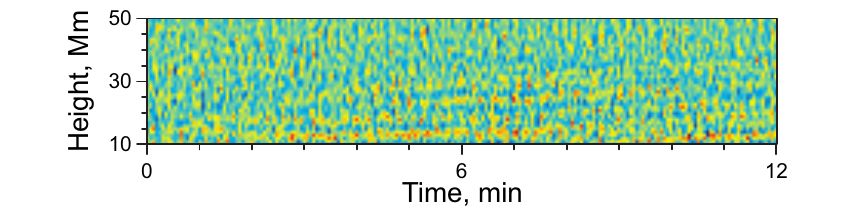}
                \label{fig:noise}
        \end{subfigure}
        
        \begin{subfigure}[]{\linewidth}
                \centering
                \caption{}
                \vspace{-6mm}
                \includegraphics{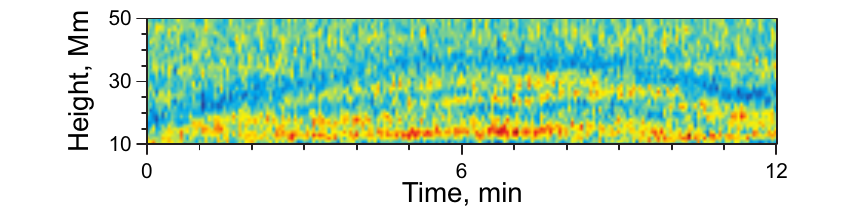}
                \label{fig:overfilt}
        \end{subfigure}

        \begin{subfigure}[]{\linewidth}
                \centering
                \caption{}
                \vspace{-6mm}
                \includegraphics{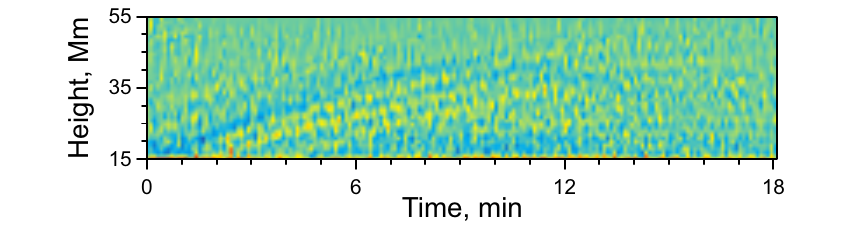}
                \label{fig:wave}
        \end{subfigure}
        \caption{High-frequency component of the density maps removed by the low-pass filter using the optimal~(\subref{fig:noise}) and excessively low cut-off frequency~(\subref{fig:overfilt}), and the train of acoustic waves revealed by this procedure in macrospicule No.~10~(\subref{fig:wave}). Blue is negative and red is positive. 
}
\label{fig:filt}
\end{figure}

At first, we smoothed away the CCD and frame-to-frame noise by applying a low-pass filter in both spatial and temporal dimensions.
The cut-off frequencies were set up manually based on the diagrams of the high-frequency component removed by the filter, which are illustrated in Fig.~\ref{fig:filt}.
For each individual macrospicule we found the minimum cut-off frequency whereby the noise remained visually uniform as shown in Fig.~\ref{fig:noise} and no distinct large- and medium-scale features appeared as in Fig.~\ref{fig:overfilt}.
This helped us to preserve as much detail on the density maps as possible while eliminating most of the sudden changes, which would result in sharp velocity outbursts.
An example of the filtered density map is shown in Fig.~\ref{fig:filtslice}.

Furthermore, being located close to the solar disc, macrospicules are observed against a comparatively strong background which is highly non-uniform in the radial direction.
This background probably originates from the thermal noise of the CCD and the scattered radiation, but mainly represents the emission of the undisturbed corona itself \citep{1995ApJ...442..653J, 1999SoPh..188..259D}.
In the context of our study, this background is equivalent to the presence of a certain amount of unmovable mass, which will perturb the calculated velocities, making them proportionally lower.
Therefore, for each macrospicule we determined the background as
\begin{equation}
B(z) = \min_{t}\lambda(z,t) 
\end{equation}
and, in order to eliminate the small-scale irregularities, we fitted this minimum profile with a smooth function
\begin{equation}
\widetilde{B}(z) = a\textrm{e}^{ -bz + c\textrm{e}^{-dz} } \,,
\label{eq:bg_model}
\end{equation}
where $a$, $b$, $c$, and $d$ are the free parameters.
This function has been found to be a good approximation to the quiet-Sun off-limb radial profiles at above 15~Mm in the \ion{He}{ii} 304~\AA\ line \citep{2015SoPh..290.1963L}.
We subtract this fitted background from the density maps, the result of which is shown in Fig.~\ref{fig:noBGslice}.

Finally, there are usually artefacts in the upper part of the density map not associated with the macrospicule itself, most probably resulting from minor concentrations of cold plasma accidentally entering the cut-out window from the outside.
They take the form of isolated blobs not connected to the map boundaries, and since our model assumes that all plasma only enters from below and moves along the axis, such artefacts violate the conservation of mass. Thus, they can substantially perturb our further results for the underlying parts of the macrospicule owing to the integration method that we use.
Most of these artefacts were eliminated based on their position on the density map,
with a smooth transition function used to avoid sharp changes of density.
Figure~\ref{fig:clearslice} shows the resulting density map which served as the input for the calculation technique described below.

\subsection{Velocity field calculation} \label{velcalc}
\begin{figure}
        \centering
        \begin{subfigure}[]{\linewidth}
                \centering
                \caption{} 
                \vspace{-6mm}
                \includegraphics{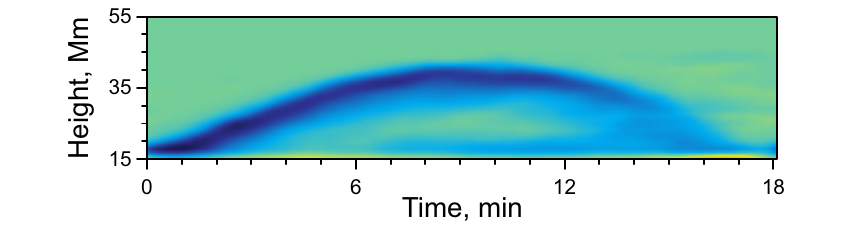}
                \label{fig:z_deriv}
        \end{subfigure} 
        
        \begin{subfigure}[]{\linewidth}
                \centering
                \caption{} 
                \vspace{-6mm}
                \includegraphics{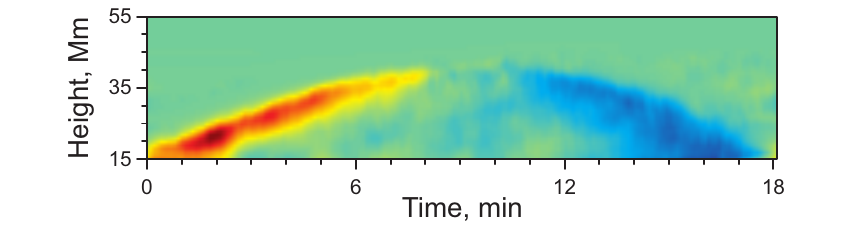}
                \label{fig:t_deriv}
        \end{subfigure}
        \caption{Normalised space~(\subref{fig:z_deriv}) and time~(\subref{fig:t_deriv}) partial derivative arrays in the $P$ and $Q$ terms. Blue is negative and red is positive.}
        \label{fig:derivs}
\end{figure}

With these estimates of how the linear plasma density changes with time along a macrospicule, we attempted to reconstruct the evolution of its velocity field based on the continuity equation.
At this stage, we assumed that loss of the observed mass as a result of both temperature change and transverse plasma flows is negligible in the sense that it does not produce essential perturbations to the calculated velocity field.
Thus, in the one-dimensional case, the continuity equation takes the form
\begin{equation}
\frac{ \partial \lambda }{ \partial t } + \frac{ \partial }{ \partial z } \left( \lambda v_z \right) = 0 \,,
\label{eq:conteq}
\end{equation}
where $v_z$ is the vertical component of the velocity vector.
In terms of finding the velocity field this yields
\begin{equation}
\frac{ \partial v_z }{ \partial z } = f(v,z) = P v_z + Q \,,
\label{eq:diffeq}
\end{equation}
where
\begin{equation}
P = -\frac{1}{ \lambda } \frac{ \partial \lambda }{ \partial z }
\label{eq:deriv_z}
\end{equation}
and
\begin{equation}
Q = -\frac{1}{ \lambda } \frac{ \partial \lambda }{ \partial t }
\label{eq:deriv_t}
\end{equation}
are data arrays of the same dimensions as the density map $\lambda$, and were calculated in advance (Fig.~\ref{fig:derivs}).
The term $ 1/\lambda $ in Eqs.~(\ref{eq:deriv_z}) and (\ref{eq:deriv_t}) makes it necessary to introduce a small offset to $\lambda$ (which is set up manually for each individual macrospicule) in order to make it non-zero at any point in space and time.

For the boundary conditions, we assumed that there should be no plasma motions at $z\rightarrow+\infty$.
In practice, this means that the velocity must be set to zero above a certain height $z_{\mathrm{max}}$, which should be high enough to ensure that none of the macro\-spicule's material reaches this height, but as low as possible in order to accumulate less error from the upper part of the density map in computation.
Typically, we set $z_\mathrm{max}$ at around 5~Mm above the maximum height of the macrospicule's elevation.

To numerically solve the differential equation~(\ref{eq:diffeq}) we employed the five-step Adams–Bashforth method.
This is an explicit linear multistep integration method, which has an advantage over {e.g.} the Runge-Kutta methods, as it is known to be of higher precision and does not require the function to be defined in the intermediate grid points \citep{hairer2011solving}.
The method used here is zero stable and has an error of the order of $O (h^5)$, where $h$ is the grid step \citep{suli2003introduction}.
With this technique, each column of the data array was integrated independently starting from its top, where the boundary condition was set in the five additional grid nodes.
We have tested this method on a number of model datasets, and the results are in good correspondence with the pre-set velocity field, except for being slightly smoothed owing to the offset introduced to $\lambda$ at the previous stage.

\section{Results} \label{sect:results}
\subsection{Velocity field} \label{sect:velfield}

\begin{figure}
        \centering
        \includegraphics{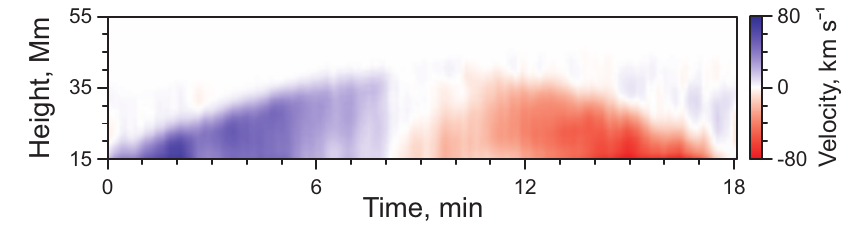}
        \caption{Reconstructed evolution of the macrospicule's one-dimensional velocity field as projected on the vertical axis. Upward velocities are shown in blue and downward in red.}
        \label{fig:velfiled}
\end{figure}

Using this method, which is still limited by the assumptions of our model, we obtained the vertical component of the plasma velocity field for 18 of the 36 macrospicules as a function of time, an example of which is shown in Fig.~\ref{fig:velfiled}.
For the remainder, no plausible results could be achieved as they were either too faint against the background or not sufficiently elevated.
For 12 of the 18 macrospicules the evolution of the velocity field shows a similar recurrent behaviour, as the macrospicule rises above the limb with a decreasing upward velocity, stops, and then retracts with an increasing downward velocity.
The other 3 of these 18 macrospicules have only been observed during the ascending or the descending phase, but probably followed the same pattern outside the observation period, and 3 more of them were substantially distorted or showed a more complex behaviour, most likely resulting from a superposition of two or more closely spaced non-simultaneous jets.

\begin{figure}[]
\centering
\includegraphics{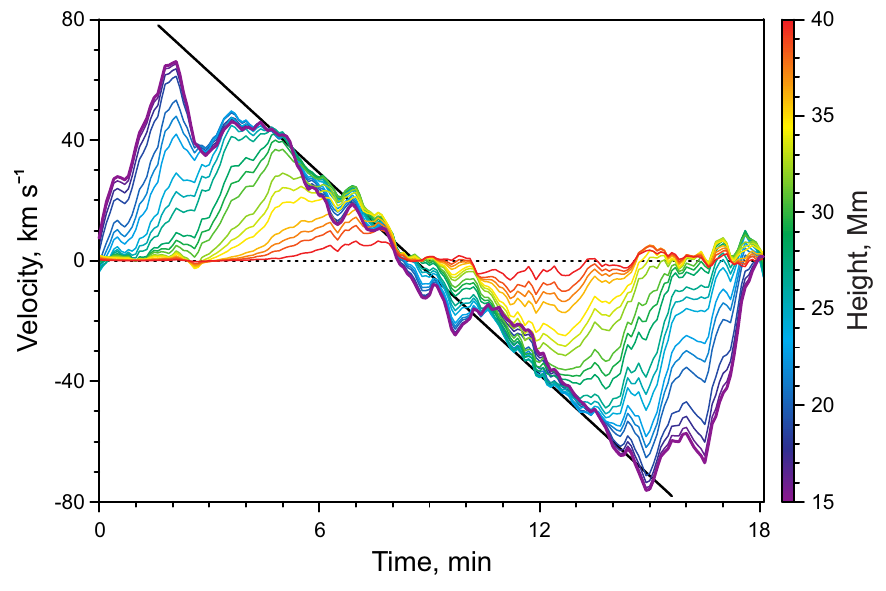}
\caption{Velocity profiles at varied heights from 15 to 40~Mm (coloured lines). The straight black line corresponds to the parabolic fit of the macrospicule's leading edge trajectory.}
\label{fig:profiles}
\end{figure}

For a more detailed analysis of a macrospicule's dynamics, velocity profiles at varying heights, such as that shown in Fig.~\ref{fig:profiles}, prove to be very informative. 
They showed that the velocity of each of the 12 macrospicules indicated above follows a similar pattern at all heights consisting of three major phases:
\vspace{-\topsep}
\begin{enumerate}[1)]
  \item As more moving plasma enters the area, the velocity increases until it reaches a certain maximum value;
  \item The velocity then starts to decrease with a nearly constant deceleration, reverses sign, and continues to decrease in the same manner;
  \item As the macrospicule's material leaves the area, the velocity finally returns to a near-zero value.
\end{enumerate}
\vspace{-\topsep}
This pattern is less prominent at greater heights reached only by the faint and diffuse top of the macrospicule ({e.g.} above 35~Mm in Fig.~\ref{fig:profiles}), where phase~2 is almost absent, and where the velocity change during phases 1 and 3 becomes noticeably flatter.
We attribute this to the fact that the macrospicule's density in this domain becomes comparable to the previously introduced offset.
Finally, the most interesting result revealed by these plots is that during phase~2 the velocity profiles are most often located very close together and are, moreover, in a good correspondence with the macrospicules' leading edge trajectories (black line in Fig.~\ref{fig:profiles}) discussed in the next section.

Beyond that, for one of the macrospicules (No.~10 in Table~\ref{tbl:spicstats}, which was chosen for illustration in Figs.~\ref{fig:slices}, \ref{fig:derivs}, \ref{fig:velfiled}, \ref{fig:profiles}, and~\ref{fig:loss}) we identified a train of three compression waves which start to propagate upwards from the macrospicule's base in the early phase of its motion.
The waves are scarcely seen in the initial density map, but are clearly identifiable in the subtracted high-frequency component (Fig.~\ref{fig:wave}) as a set of fringes.
They arise mainly from intensity variations, and we are inclined to interpret them as the magneto-acoustic waves since the intensity change is more likely to result from the density oscillations, rather than from a change in temperature or in ionisation degree.
The inclination of the fringes and the distance between them yields a varying propagation velocity, which gradually decreases from $52 \pm 7$~km~s$^{-1}$ for the first wave to $42 \pm 5$~km~s$^{-1}$ for the last one, and a wave period of roughly $180 \pm 30$~s.
Most probably, the leftovers of these waves in the filtered density map are responsible for the quasi-periodic disturbances in the velocity field, seen in Fig.~\ref{fig:profiles} at $t = 3$, $t = 6.5$, and $t = 9.5$~min.

\subsection{Leading edge trajectories} \label{sect:decel}

Along with the reconstruction of the velocity field, we studied trajectories of the apparent end points of the same 18 macrospicules, and for 15 of them we found these points to follow parabolic trajectories with high precision (Fig.~\ref{fig:clearslice}).
The obtained trajectory parameters corrected for the inclination angle are presented in Table~\ref{tbl:spicstats}.
The decelerations typically range from 160 to 230~m~s$^{-2}$ and initial velocities (defined as the extrapolated velocity at zero height) from 80 to 130~km~s$^{-1}$.
Two macrospicules, however (Nos.~2 and~4), have significantly higher decelerations, which are even larger than gravity on the solar surface $g=274$~m~s$^{-2}$, and correspondingly have significantly higher initial velocities.
Also, the maximum heights reached by the macrospicules are typically from 30 to 45~Mm, and the lifetimes are in the range of 10--20~min, which is in a good agreement with earlier studies \citep{1976ApJ...203..528W, 1989SoPh..119...55D, 1994ApJ...431L..59K}.
Again, there are two exceptions (Nos.~14 and 18), which stand out for their greater heights and longer lifetimes.
They do not follow parabolic trajectories, and probably should be classified as small surges rather than macrospicules.

\begin{table*}
        \caption{Obtained parameters of the macrospicules.} 
        \label{tbl:spicstats}
        \centering
        \begin{tabular}{c c c c c c c}
                \hline\hline  
                \noalign{\smallskip}
                No.     &       Lifetime, min   &       Height, Mm      &       Deceleration, m~s$^{-2}$      &       Ballistic deceleration, m~s$^{-2}$      &       Velocity, km~s$^{-1}$     &       Loss of mass, \%        \\
                \hline
                \noalign{\smallskip}
                1       &       16.4    &       44.3    &       181.1   &       267.0   &       125.1   &       52.2    \\
                2       &       8.4     &       34.7    &       566.7   &       265.1   &       195.2   &       9.7         \\
                3       &       15.2    &       33.1    &       183.7   &       272.3   &       110.0   &       16.7    \\
                4       &       12.8    &       35.7    &       338.3   &       270.0   &       154.2   &       15.3    \\
                5       &       11.8    &       30.9    &       228.1   &       267.9   &       117.4   &       34.7    \\
                6       &       6.9     &       26.7    &       ---     &       ---         &       ---     &       ---     \\
                7       &       >8.8    &       40.4    &       126.6   &       261.8   &       98.9    &       ---         \\
                8       &       >8.8    &       40.2    &       198.5   &       259.7   &       122.9   &       ---         \\
                9       &       11.8    &       26.8    &       159.3   &       252.5   &       88.7    &       55.8    \\
                10      &       18.1    &       41.2    &       214.5   &       250.3   &       127.0   &       46.6    \\
                11      &       19.1    &       45.0    &       224.7   &       249.1   &       131.1   &       29.7    \\
                12      &       12.0    &       28.9    &       190.9   &       274.0   &       105.0   &       16.9    \\
                13      &       20.0    &       30.8    &       170.9   &       273.3   &       102.5   &       29.3    \\
                14      &       28.0    &       59.3    &       ---     &       ---         &       ---     &       ---     \\
                15      &       >10.9   &       25.2    &       167.1   &       245.4   &       86.8    &       ---         \\
                16      &       16.5    &       23.9    &       165.9   &       222.4   &       80.2    &       79.6    \\
                17      &       12.4    &       26.4    &       264.8   &       245.8   &       112.0   &       28.4    \\
                18      &       28.0    &       61.3    &       ---     &       ---         &       ---     &       ---     \\
                \hline
        \end{tabular}
        \tablefoot{Here, deceleration is the actual value measured from the parabolic fit, while ballistic deceleration is calculated based on the Equation~(\ref{eq:grav_decel}).}
\end{table*}

Given such parabolic trajectories, it would be most natural to assume that dynamics of the macrospicules are purely ballistic and that the observed decelerations can be accounted for by the inclination of the open magnetic field lines.
In this scenario, the deceleration of the macrospicule's plasma is defined as
\begin{equation}
\label{eq:grav_decel}
a_\textrm{b} = g \cos\theta \,,
\end{equation}
where $\theta$ is the inclination angle.
Assuming that at solar minimum the open magnetic field lines guiding the macrospicules are close to being axially symmetric relative to the solar axis, it can be argued that the observed inclinations are almost free from line-of-sight effects, since the macrospicules that are significantly deflected from the image plane should be located at such latitudes where they cannot be observed off-limb because of their small heights.
Having taken this into account, we found, however, no noticeable correlation between the deceleration of macrospicules and their apparent inclination, with the Pearson's correlation coefficient being as low as~0.17.
Moreover, this ballistic scenario is not able to explain decelerations larger than $g$.

\begin{figure}[]
        \centering
        \includegraphics{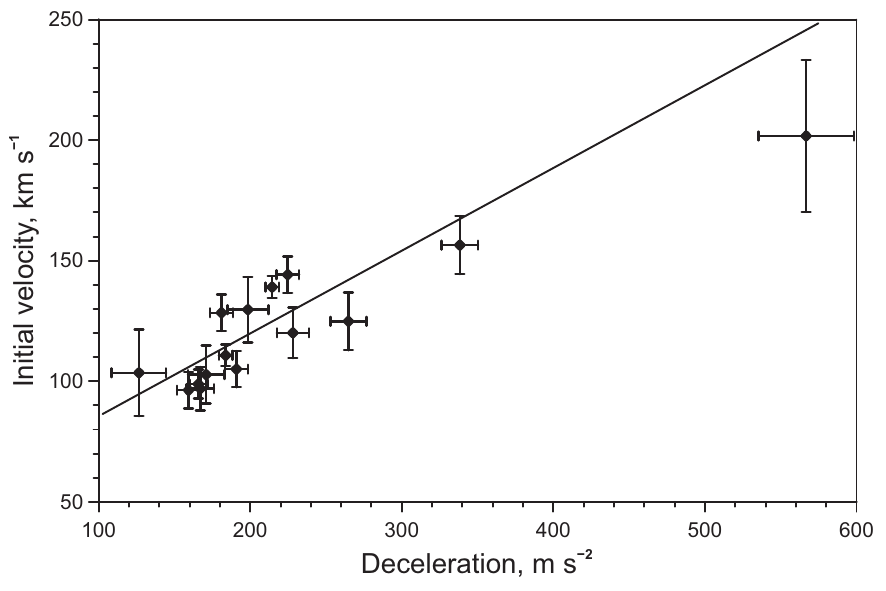}
        \caption{Decelerations versus initial velocities of the macrospicules obtained through the parabolic fits.}
        \label{fig:decel}
\end{figure}

Several other types of solar jets, such as quiet-Sun mottles, active region fibrils, and spicules are known to follow parabolic but not ballistic trajectories, {i.e.} their decelerations cannot be solely explained by solar gravity.
It was recently found that all these types of jets show a clear linear correlation between the deceleration and the maximum velocity \citep{2006ApJ...647L..73H, 2007ApJ...655..624D, 2007PASJ...59S.655D, 2007ApJ...660L.169R}, while the same dependence was reproduced in numerical simulations within the model of magneto-acoustic shock wave driven jets \citep{2007ApJ...666.1277H}.
To check whether this applies to the macrospicules as well, we plotted the decelerations $a$ versus the initial velocities $v_0$ of the 15 macrospicules studied (Fig.~\ref{fig:decel}).
We found a similar linear dependence of the form
\begin{equation}
v_0 = ka + c \,,
\end{equation}
with the coefficients $k=340 \pm 10$~s and $c=51.3\pm 1.9$~km~s$^{-1}$, and the correlation coefficient $r=0.85$.
However, the obtained coefficients $k$ and $c$ are roughly three times higher than those we inferred from the plots in papers cited above.
This suggests that even if the macrospicules are driven by the magneto-acoustic shocks as well, at least some of their formation conditions are likely to be different.

\subsection{Loss of mass estimates}

In previous analysis we had to neglect possible losses in order to leave $v_z$ as the only unknown variable in the continuity equation~(\ref{eq:conteq}).
However, a reasonably good correspondence between the calculated velocity profiles and the fitted parabolic trajectories allows us to overcome this complication.
As illustrated by Fig.~\ref{fig:profiles}, in some cases the plasma velocity in the lowest observable part of a macrospicule (thick purple line) for almost all of its lifetime coincides with the velocity of the leading edge $\overline{v}_z$ (black line) obtained through the parabolic fit.
The same applies for the overlying velocity profiles, but to a lesser degree since phase~2 takes up a smaller part of the motion pattern at those heights.

This leads us to the idea that plasma velocity can be constant along the macrospicule's axis and equal to $\overline{v}_z$ at every fixed moment of time.
Therefore, we can introduce a new variable, the loss rate $-\sigma$, to the continuity equation, which will take the form
\begin{equation}
\frac{ \partial \lambda }{ \partial t } + \overline{v}_z \frac{ \partial \lambda}{ \partial z } = \sigma \,.
\label{eq:cont_loss}
\end{equation}
This assumption allows us to invert the problem and obtain rough estimates of the amount of visible mass lost by a macrospicule throughout its presence in the lower corona.
We further assume that the loss rate is uniform in time and space, thus being responsible for a certain systematic discrepancy in the previously calculated velocity field.
It can be clearly seen that there is still a significant number of more random fluctuations, which we attribute to the inaccuracy of our model and external effects.
We suppose, however, that owing to its randomness, this factor will only result in a wider distribution of the obtained $\sigma$ values, while not affecting the average.

\begin{figure}[]
        \centering
        \includegraphics{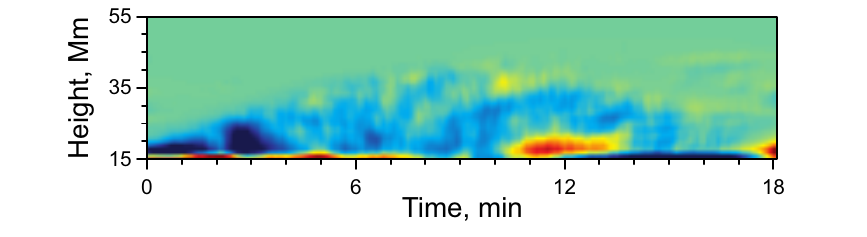}
        \caption{Space-time distribution of the loss rate $-\sigma$. Mass inflow is shown in red and mass outflow in blue.}
        \label{fig:loss}
\end{figure}

Equation~(\ref{eq:cont_loss}) leaves the calculation of $\sigma$ rather straightforward; having found it as a function of height and time (Fig.~\ref{fig:loss}), we calculated the total mass loss of the macrospicule as
\begin{equation}
L_{\mathrm{\Sigma}} = \int\displaylimits_{t_1}^{t_2} \int\displaylimits_{z_1}^{z_2}\! - \sigma(z,t)\ dz\ dt  \,,
\end{equation}
where $t_1$ and $t_2$ are the time stamps of the macrospicule's appearance and disappearance above the limb, respectively; $z_1$ is the minimum height where the macrospicule becomes distinguishable from the background (typically 10--15~Mm); and $z_2$ is the maximum height reached by the macrospicule.

Since plasma densities were previously determined in the arbitrary units, we are only able to obtain the relative loss
\begin{equation}
\widetilde L = \frac{L_{\mathrm{\Sigma}}}{M_{\mathrm{\Sigma}}} \,,
\label{lossrate}
\end{equation}
where $M_{\mathrm{\Sigma}}$ is the total mass of the macrospicule, which is defined as the maximum amount of the mass visible off-limb when the macrospicule is maximally elevated, i.e.
\begin{equation}
M_{\mathrm{\Sigma}} = \max_{t} \int\displaylimits_{z_1}^{z_2} \! \lambda(z,t)\ dz \,.
\end{equation}

Following this procedure, we calculated the loss of mass percentage for 12 of the macrospicules studied.
We found that most of them lose from 10 to 30\% of their visible mass, and four macrospicules lose an even greater amount --- three around 50\% and one almost 80\% --- with a strong fading being noticeable in the density maps in all four cases.
We should stress, however, that these are only rough estimates, and therefore must be treated with caution.
The obtained $\sigma$ values are highly non-uniform and often have significant perturbations, such as those at $t=3$~min and $t=12$~min in Fig.~\ref{fig:loss}, especially at the lower heights.
Most probably they result from external plasma flows not associated with the macrospicule itself entering the cut-out window either from the sides ({e.g.} neighbouring macrospicules) or from below (EUV spicules), as well as from other processes, such as the magneto-acoustic waves discussed in Section~\ref{sect:velfield}.
Consequently, the obtained $\sigma$ values have relatively wide distributions, which are, nevertheless, always centred in the negative range. 
From these distributions we estimated the relative error of the latter results as around 50--60\%.

With the typical electron density of $10^{10}$~cm$^{-3}$ \citep{1976ApJ...203..528W, 1991ApJ...376L..25H}, hydrogen abundance and ionisation degree close to~1 (for simplicity as the temperature is close to or above  $10^4$~K), diameter of 5~Mm, length of 30~Mm \citep{1976ApJ...203..528W, 1989SoPh..119...55D}, the observed birth rate of around $2 \times 10^{-2}$~s$^{-1}$ in a single coronal hole (this observation), and an estimate that only around $\nicefrac{1}{2}$ of all the macrospicules are visible off-limb (for the estimation technique see {e.g.} \citealp{1998ApJ...509..461W, 2015SoPh..290.1963L}), the inferred loss rate amounts to roughly $10^8$~kg~s$^{-1}$ of plasma, which is presumably heated to higher temperatures, thus supplying mass into the corona and, consequently, to the solar wind.
This value constitutes around 5--10\% of the total solar wind flux observed in the Earth's orbit, which makes macrospicules a non-negligible factor in the mass balance of the solar atmosphere.

\section{Discussion} \label{sect:discussion}

High-cadence observations have a great advantage of allowing hydrodynamic methods of investigation and thus of studying the internal motions of a highly dynamic object in great detail.
Using such data from the TESIS space project with a cadence of 3.5--6.0~s we reconstructed the evolution of the one-dimensional velocity field for several macrospicules.
We established a strong linear correlation between a macrospicule's deceleration and initial velocity, which together with the magneto-acoustic waves observed in one of the macrospicules gives a direct indication towards the mechanism behind their parabolic motion.
We also found a close correspondence between the internal velocities and the parabolic trajectories of the macrospicules' leading edges, which, in turn, allowed us to introduce a loss factor into the calculations and give an early estimate of the typical amount of visible mass which dissipates during a macrospicule's movement in the low corona.

Such estimates are essential for our understanding of the role of macrospicules in the mass balance of the corona and in the formation of the fast solar wind component.
When a relatively cool structure such as a macrospicule moves through the hot corona, the dissipation of its visible mass is most likely to result from the heating of some portion of its material, thus giving out new plasma into the corona.
We must note, however, that in the above we have implicitly assumed that all of the macrospicule's material is initially fully visible in \ion{He}{ii} 304~\AA\ and then gradually fades out of the line.
Though highly probable since macrospicules are best seen in the \ion{He}{ii} 304~\AA\ line, this is not necessary true as a certain amount of cooler plasma could be heated to the transition region temperatures during a macrospicule's movement.
Alternatively, as macrospicules are moving up, their material can expand and cool down, thus fading out of the \ion{He}{ii} line but not supplying plasma to the corona.
These two factors can either under- or overestimate the calculated $\sigma$ values, and thereby introduce further uncertainty to what portion of the macrospicule's mass actually ends up in the corona.

The main source of error, though, is still the optical thickness of macrospicules in the \ion{He}{ii} 304~\AA\ line.
This allows only the outer sheath of a macrospicule to be effectively observed, while giving us no precise information about plasma motions in its interior.
Therefore, to apply our method we had to assume a fairly simple radiative model and the symmetric inner structure of a macrospicule, which had the greatest impact on our estimate of the loss rate.
We believe, however, that the assumption of axial symmetry is a reasonable one in the case of macrospicules.
A possible solution to this issue would be to study macrospicules in spectral lines where they appear to be optically thin (showing up in either emission or absorption) and to directly obtain full column densities.
In this respect a notable advantage of macrospicules is that individual features can be easily isolated.

Moreover, we have obtained the velocity field on the assumption of negligible loss, which is in  contradiction with the obtained estimates of mass dissipation.
Nevertheless, it was supposed in the above that the loss of mass is nearly uniform, and therefore, that it should not disrupt the calculated velocity field significantly.
In this context, our assumption of the whole macrospicule moving with the same velocity as its end point still appears to be plausible.
Another limitation of the technique applied is that it gives only plasma velocities in the image plane.
This could be surpassed with the help of spectroscopic observations, such as those performed by \citet{2005A&A...438.1115X}, which will yield the line-of-sight velocities necessary for the reconstruction of the three-dimensional velocity vector.
However, as we note in Section~\ref{sect:decel}, the observed macrospicules are not likely to be very much deflected from the image plane.

Finally, in this study we examined only two short time series and obtained full results for only 12 macrospicules of the 36 detected.
It would therefore be highly advisable to make use of other available data sources.
The most promising perhaps is the vast pool of the Solar Dynamics Observatory (SDO) data, which have a lower cadence but a higher angular resolution, and more importantly, make up a continuous observation.
We are now working to apply this technique to effectively process vast amounts of the SDO data with a minimum of manual operations.
This will allow us to perform a statistical analysis of their properties and to detect possible variations throughout the solar cycle, and also to give a better account of their global impact on the mass balance in the corona.
A major obstacle in this endeavour is that to date SDO has never operated during the solar minimum, which results in a more complex geometry of the coronal holes where the macrospicules are usually observed.


\begin{acknowledgements} 
This work was supported by the Russian Foundation for \mbox{Basic} Research (grant 14-02-00945) and by  Programme No.~9 for fundamental research of the Praesidium of the Russian Academy of Sciences.
We are grateful to all members of the TESIS development team who have put much effort into the design and operation of this unique instrument.
We would also like to thank Antoine Reva for his early insight into the subject and for fruitful discussions.
\end{acknowledgements}

\bibliography{macrospic}

\end{document}